\documentclass{PoS}
\pdfoutput=1

\newcommand{\GeV}{\textrm{GeV}}

\newcommand{\be}{\begin{eqnarray}}
\newcommand{\ee}{\end{eqnarray}}

\renewcommand{\sc}{\slashchar}
\newcommand{\ds}[1]{
#1{\hskip-2.0mm}/
}
\def\slashchar#1{\setbox0=\hbox{$#1$}           
   \dimen0=\wd0                                 
   \setbox1=\hbox{/} \dimen1=\wd1               
   \ifdim\dimen0>\dimen1                        
      \rlap{\hbox to \dimen0{\hfil/\hfil}}      
      #1                                        
   \else                                        
      \rlap{\hbox to \dimen1{\hfil$#1$\hfil}}   
      /                                         
   \fi}                                         %

\title{Exploring the transition into the Chiral Regime of QCD using the Interacting Instanton Liquid Model}

\ShortTitle{Exploring the Chiral Regime of QCD in the Interacting Instanton Liquid Model}

\author{\speaker{Marco Cristoforetti}\\ 
        Trento University and I.N.F.N. Gruppo Collegato di Trento, Via Sommarive 14 Povo (Trento), 38100 Italy\\
        E-mail: \email{cristof@science.unitn.it}}
\author{Pietro Faccioli  \\
        Trento University and I.N.F.N. Gruppo Collegato di Trento, Via Sommarive 14 Povo (Trento), 38100 Italy\\
        E-mail: \email{faccioli@science.unitn.it}}
\author{Marco Traini\\
        Trento University and I.N.F.N. Gruppo Collegato di Trento, Via Sommarive 14 Povo (Trento), 38100 Italy\\
        E-mail: \email{traini@science.unitn.it}}
\author{John W. Negele,\\ 
				Center for Theoretical Physics Massachusetts Institute of Technology, NE25-4079, 
77 Massachusetts Ave, Cambridge, MA 02139-4307, USA. \\
				E-mail: \email{negele@lns.mit.edu}}
\abstract{The non-perturbative  quark-gluon interaction depends significantly on the value of the quark mass. In particular, in the light quark mass regime, correlations are strongly influenced by dynamics associated to chiral symmetry breaking. We use the Interacting Instanton Liquid Model (IILM) as a tool to investigate the microscopic dynamical mechanisms which underly the dependence on the quark mass and drive the transition into the chiral regime of QCD. To ensure the validity of the model, we first verify that the dependence on the quark mass for several observables calculated in the IILM agrees well with the predictions of chiral perturbation theory and with lattice simulations. We then show that a quark mass $m*\approx80$ MeV emerging naturally from the model specifies the mass scale above which the dynamics associated with low-lying eigenmodes of the Direac operator becomes sub-leading and the contribution of the fermion determinant is suppressed.}

\FullConference{The XXV International Symposium on Lattice Field Theory\\
		 July 30-4 August 2007\\
		 Regensburg, Germany}

\begin{document}

\section{Introduction}\label{intro}
Although Lattice QCD provides the only available ab-initio tool for solving QCD non-perturbatively, such a framework does not allow a direct identification of the dynamical mechanism underlying hadron structure and of the relevant degrees of freedom at different scales. In this talk we present our recent attempt to extract complementary insight using the Interacting Instanton Liquid Model (IILM)~\cite{chptiilm}.

In the IILM, the QCD path integral over all possible gluon field configurations is replaced by an effective theory in which instantons are the effective degrees of freedom, and the gauge fields of the theory are those generated by integrating over the positions, color orientations, and sizes of instantons. The QCD partition function reduced to:
\begin{equation}
\label{eq:partf}\nonumber
\mathcal{Z}_{QCD}\simeq\mathcal{Z}_{ILM} =\sum_{N_+, N_-} \frac{1}{N_+!N_-!}\int\prod_i^{N_++N_-}\textrm{d}\Omega_id(\rho_i)e^{-S_{int}}\prod_i^{N_f}\textrm{det}(i\sc{D}+im_f).
\end{equation}
Here, $\textrm{d}\Omega_i=\textrm{d}U_i\textrm{d}^4z_i\textrm{d}\rho_i$ is the measure in the space of collective coordinates, color orientation, position and size, associated with the single instantons. Quantum fluctuations are included in Gaussian approximation, through the semi-classical instanton amplitude $d(\rho_i)$. $S_{int}$ is a bosonic  interaction between pseudo-particles which includes a phenomenological short-range repulsive core required to remove large-sized instantons from the vacuum. In the formulation of the model we have considered, the strength of such a repulsion is the only phenomenological parameter, which has to be tuned to reproduce observations.

Numerical simulations in the Instanton Liquid Model performed in the last few years have been shown to be in very good agreement with both phenomenology \cite{emff3,delta12} and lattice simulations \cite{latticeILM3}. Here, we would like to understand the mass range in which instanton mediated chiral dynamics is manifest and the extent to which it is described by chiral perturbation theory.
In particular, we would like to identify the mass scale or scales above which the contribution of the  fermion determinantis suppressed, and the near-zero modes of the Dirac operator become sub-dominant.  

\section{The Chiral regime in the IILM}
\begin{figure}
\centering
\includegraphics[width=.6\textwidth]{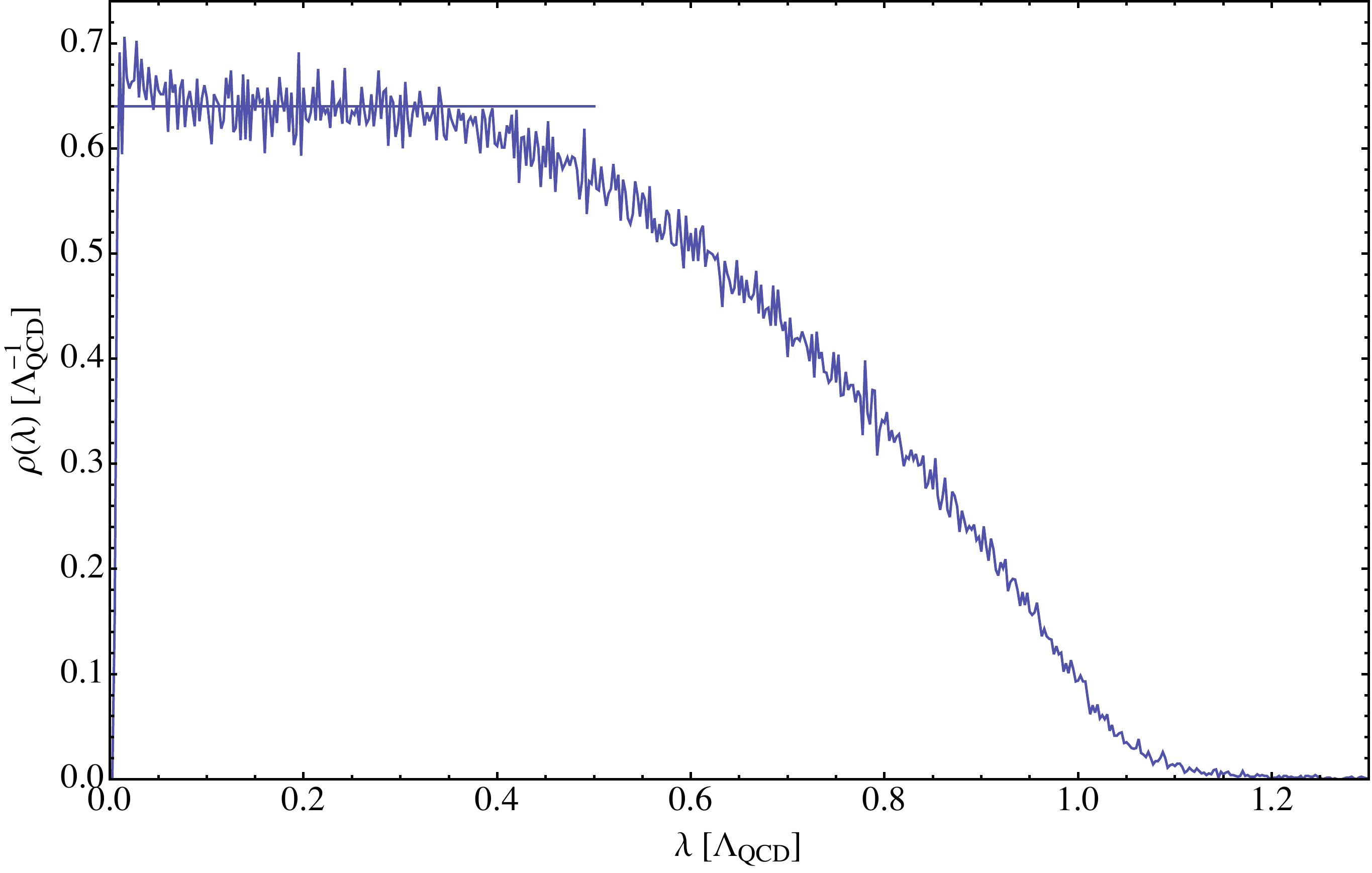}
	\caption{ Spectral density of the Dirac operator for $N_c=3$, $N_f=2$ and $m_q=0$.}
	\label{fig:spectr}
\end{figure}

In order to investigate hadron structure in the chiral regime of QCD we first check that IILM gives predictions in agreement with $\chi$pt. To test the IILM in the chiral regime, we have analyzed the structure of the spectrum of the Dirac operator. In $\chi$pt there are clear statements on how the distribution of eigenvalues of the Dirac operator $\rho(\lambda)$ changes with  the mass of current quarks. In the chiral limit, $m_q=0$, the Banks-Casher relation implies that $\rho(\lambda)$ becomes flat at the origin~\cite{smilga}
\be
\lim_{\lambda\to 0}\lim_{m_q\to 0} \rho(\lambda)= \textrm{Const} + O(\lambda^2)\qquad (N_f=2),
\ee
In Fig.~\ref{fig:spectr} we present the Dirac spectrum computed in the IILM for $m_q=0$ and $N_f=2$. Clearly, in the  $\lambda\to 0$ limit, we observe that the spectrum develops the expected plateau.

If the quark mass is small but finite, $\chi$pt predicts a peak near the origin of the form $\rho(\lambda)-\rho(0)\propto m^{\alpha}\lambda^{1-\alpha}$. Such a particular structure of the Dirac spectrum away from the chiral limit is reflected in a specific dependence of the quark condensate on the quark mass:
\be
	\langle\overline{q}q\rangle=-2m_q\int_0^{\infty}\textrm{d}\lambda\frac{\rho(\lambda)}{m_q^2+\lambda^2}
\ee
Using this relationship, we can obtain the IILM prediction for  the condensate for different $m_q$, which can be compared with the $\mathcal{O}(p^4)$ $\chi$pt formula
\be\label{eq:qqchpt}
	\langle0|\overline{q}q|0\rangle=B_0f_0^2\Big(1-\frac{6B_0m_q}{32\pi^2f_0^2}\ln\Big(\frac{2m_qB_0}{\Lambda^2}\Big)\Big).
\ee
We have found that  the observed quark mass dependence is consistent with the form expected from chiral perturbation theory. In addition, the low-energy constants $f_0$, $B_0$ and $\overline{l}_3$ extracted from a best-fit of the IILM data using (\ref{eq:qqchpt}) are consistent with the value extracted from phenomenology and recent lattice results \cite{ETM}:
\be
	f_0=0.085\pm0.005, & B_0=2.4\pm0.2, & \overline{l}_3=3.80\pm0.04,
\ee
which give $\langle0|\overline{q}q|0\rangle_0=(0.26\pm0.01\ \textrm{GeV})^3$.

\begin{figure}
\center
\includegraphics[width=.6\textwidth]{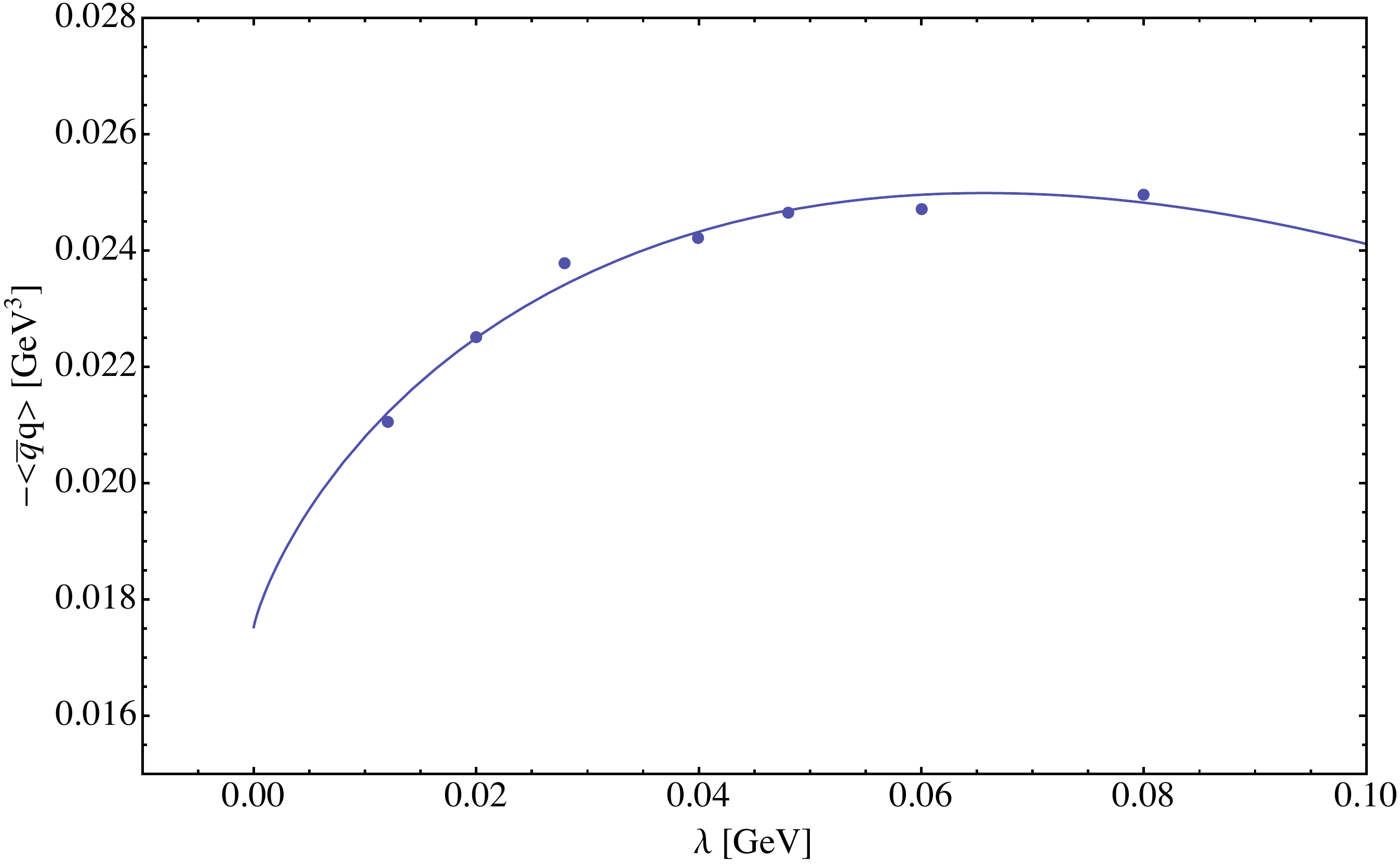}
	\caption{ Dependence of the quark condensate from the quark mass, the fit is obtained using the~(2.3).}
	\label{fig:qqfit}
\end{figure}

\section{Pion and Nucleon mass in the IILM}
The second step in our investigation of the instanton contribution to the low-energy sector of QCD is to verify that the IILM  can be used to calculate hadron masses and that these masses depend on the quark mass in the specific way predicted  from $\chi$pt. We have considered pion and nucleon because these are the two hadrons in which the instanton-induced interaction is most intense. To calculate the hadron masses we have adopted standard effective mass plot analysis, which is routinely used to extract lowest-lying hadron masses from lattice simulations.

To compute the mass of pion and nucleon we have evaluated the correlation function:
\be
G_{\pi}(\tau)=\int\textrm{d}^3\underline{x}\langle0|T[j_5(\underline{x},\tau)\overline{j}_5(\underline{0},0)|0\rangle,\qquad
G_N(\tau)=\int\textrm{d}^3\underline{x}\langle0|T[j_N(\underline{x},\tau)\overline{j}_N(\underline{0},0)P_+|0\rangle,\nonumber
\ee
where the two currents are given by $j_{N}^a(x)=\varepsilon^{abc}~u^a(x) C\gamma_5 d^b(x) u^c(x)$ and $j_{5}^a(x)=\overline{q}(x)\gamma_5\tau^a q(x)$, and $P_+=\frac{1-\gamma_4}{2}$ is the positive-parity projector.

The mass of these hadrons can then be extracted from the plateau in the large Euclidean time limit of the effective mass, i.e. using
\be
M_{\pi/N}=\lim_{\tau\rightarrow\infty}M_{\pi/N}^{eff}(\tau),&&M_{\pi/N}^{eff}(\tau)=\frac{1}{\Delta \tau}\ln\frac{G_{\pi/N}(\tau)}{G_{\pi/N}(\tau+\Delta\tau)}.\nonumber
\ee

We have computed these correlation functions for five different values of the quark mass in the range $20\lesssim m_q\lesssim 90$ MeV, finding the results reported in Table~\ref{table:hadr}.  We have found that all the pion effective mass plots display a very clean plateau, from which it is possible to unambiguously  read-off the pion mass, from a correlated chi-square fit. In the interpolation of the effective mass plots of the nucleon not only statistical but also systematic error has to be taken into account. 
\begin{table}[h!]
\caption{Pion and nucleon masses fitted by effective mass plot with $\chi^2/ndf\leq1$. The quark masses are determined at a scale $2\ \GeV$ as in \cite{musakh,bowman}}
\begin{center}
\begin{tabular}{lll}\hline\hline
	\rule{0pt}{3ex} $m_q$ [GeV]& Pion [GeV]&Nucleon [GeV]\\ \hline
  \rule{0pt}{3ex}
	$0.021 $&$ 0.300\pm 0.004 $&$ 1.11^{+0.05}_{-0.05} $\\
	$0.03 $&$ 0.360\pm 0.004 $&$ 1.15_{-0.07}^{+0.01} $\\
	$0.05  $&$ 0.460\pm 0.004 $&$ 1.20_{-0.02}^{+0.02} $\\
	$0.07 $&$ 0.530\pm 0.004 $&$ 1.28_{-0.01}^{+0.01} $\\
	$0.09  $&$ 0.600\pm 0.004 $&$ 1.35_{-0.1}^{+0.02} $\\
\hline\hline
\end{tabular}
\end{center}
\label{table:hadr}
\end{table}

\begin{figure}
	\centering
  	\includegraphics[width=.8\textwidth]{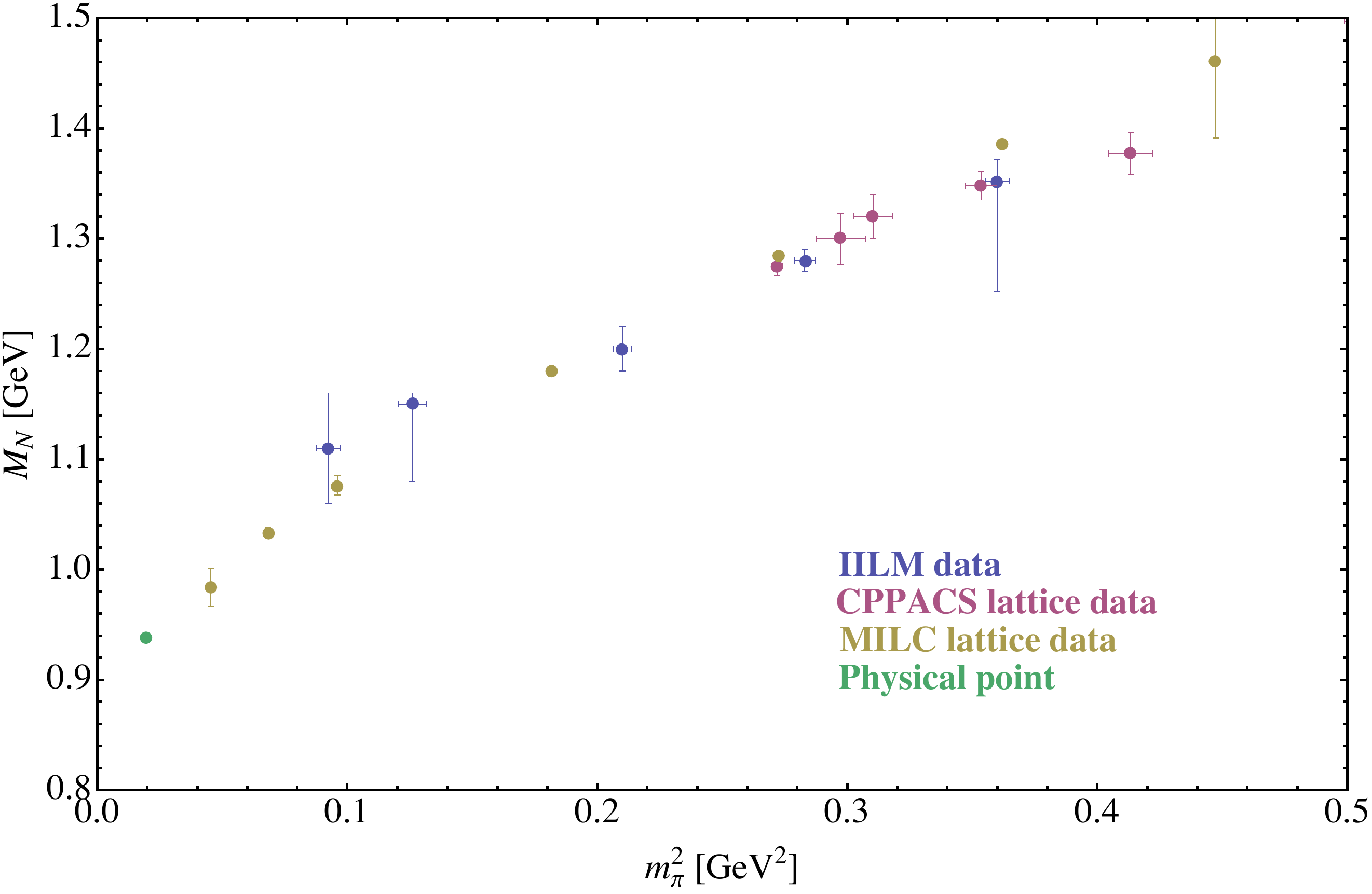}
  \caption{The nucleon mass as a function of the pion mass squared obtained in IILM and in lattice calculations. The blue points correspond to the IILM results, the yellow points to CP-PACS lattice result \cite{latticed} and the red points to MILC data \cite{latticeMILC}. Since MILC and CP-PACS adopt different scheme for fixing the physical units, we have rescaled the lattice spacing of MILC data until the two set of lattice data are consistent.  }\label{fig:latt}
\end{figure}

In order to assess the accuracy of the IILM prediction for the nucleon at different quark masses, we have verified that these are consistent with the corresponding unquenched \cite{latticed,latticeMILC} lattice QCD calculation (see Fig.~\ref{fig:latt}). 

\section{Instanton-induced dynamics in the chiral regime}
Having verified the consistency of the IILM  with lattice simulations and with the behavior expected from $\chi$pt, we conclude that restricting to semi-classical vacuum fluctuations indeed represents a useful approximation to the QCD path integral. Hence, we can use this model to explore the dynamical mechanisms controlling the role of chiral dynamics at different mass scales.  An essential  quantity in the IILM is the overlap matrix $T_{ij}=\int d^4 z \psi^{0\,\dagger}_I(z) i D_\mu \gamma_\mu \psi_j^0(z)$ specifing the probability that a quark hops from instanton $i$ to instanton $j$, this matrix plays the main role both in the evaluation of the fermionic determinant and of the quark propagator. In the IILM these two quantities can be factorized in a zero mode part and a non-zero mode part
\be
&&\textrm{Det}_A(\ds{D}+m_q)=\textrm{Det}_{A\,zm} \times \textrm{Det}^\prime_{A\,nzm},\nonumber\\
&&S(x,y)_A=S_{zm}(x,y)_{A}+S_{nzm}(x,y)_{A}.\nonumber
\ee 
where
\be
\textrm{Det}_{A\,zm}&=&\textrm{Det}(T+m_q),\nonumber\\
S_{zm}(x,y)_A&=&\sum_{I J} \psi_I^{0}(x)\psi_J^{0\dagger}(y)~\left[\frac{1}{\hat{T}-i m_q}\right]_{I J}.\nonumber
\ee
The non-zero mode part of the determinant is approximated with the product of the non-zero mode contributions of each individual instanton:
\be
\textrm{Det}^\prime_{A\,zm}=\prod_{i}^{N_++N_-}(1.34 m_q \rho_i).
\label{detnzm}
\ee
The non-zero part of the propagator is known analytically and is found to give sub-leading contribution to low-energy dynamics. 
\begin{figure}
	\centering
\includegraphics[width=0.6\textwidth]{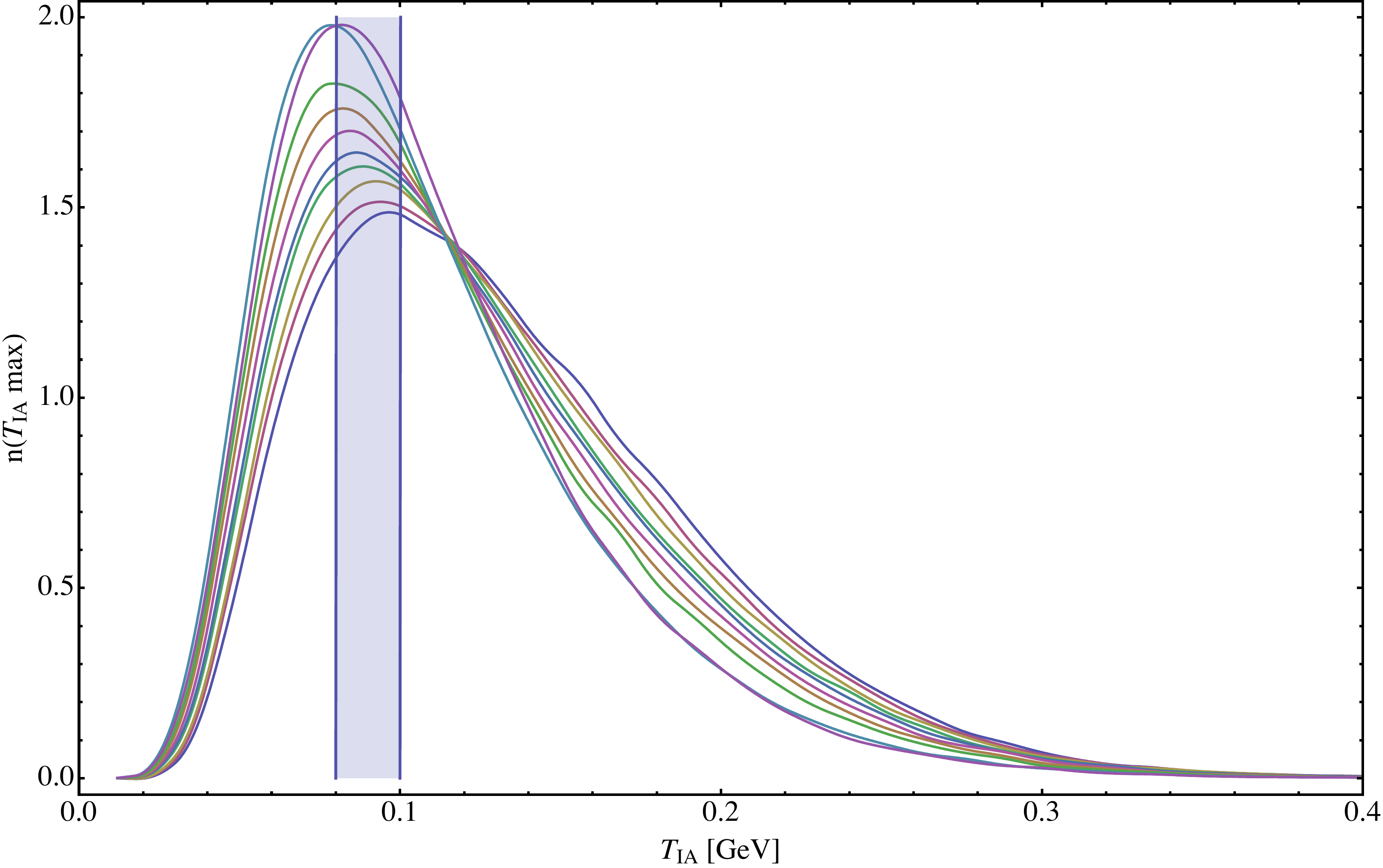}
\caption{Distribution of the maximum fermionic overlap matrix elements in the IILM for different quark masses.}
\label{fig:tij}
\end{figure}
From the structure of the quark propagator and of the fermionic determinant in the IILM model, it follows immediately that the role of instanton-induced chiral dynamics depends crucially on the value of the typical $T_{ij}$ matrix element. In Fig.~\ref{fig:tij} we have plotted the distributions of the {\it maximum} overlap matrix elements $T_{IA}$ obtained at different values of the quark mass. We note that such distributions are peaked around the value $m^\star\simeq~80$~MeV. The value $m^\star$ plays a central role in specifying the scale for instanton induced chiral symmetry breaking in the IILM. In fact, for quark masses $m_q\gg \textrm{max}[T_{IJ}]\simeq m^\star$ the contribution of the overlap matrix elements to the propagator and to the fermionic determinant becomes negligible.

\begin{figure}
	\centering
\includegraphics[width=0.55\textwidth]{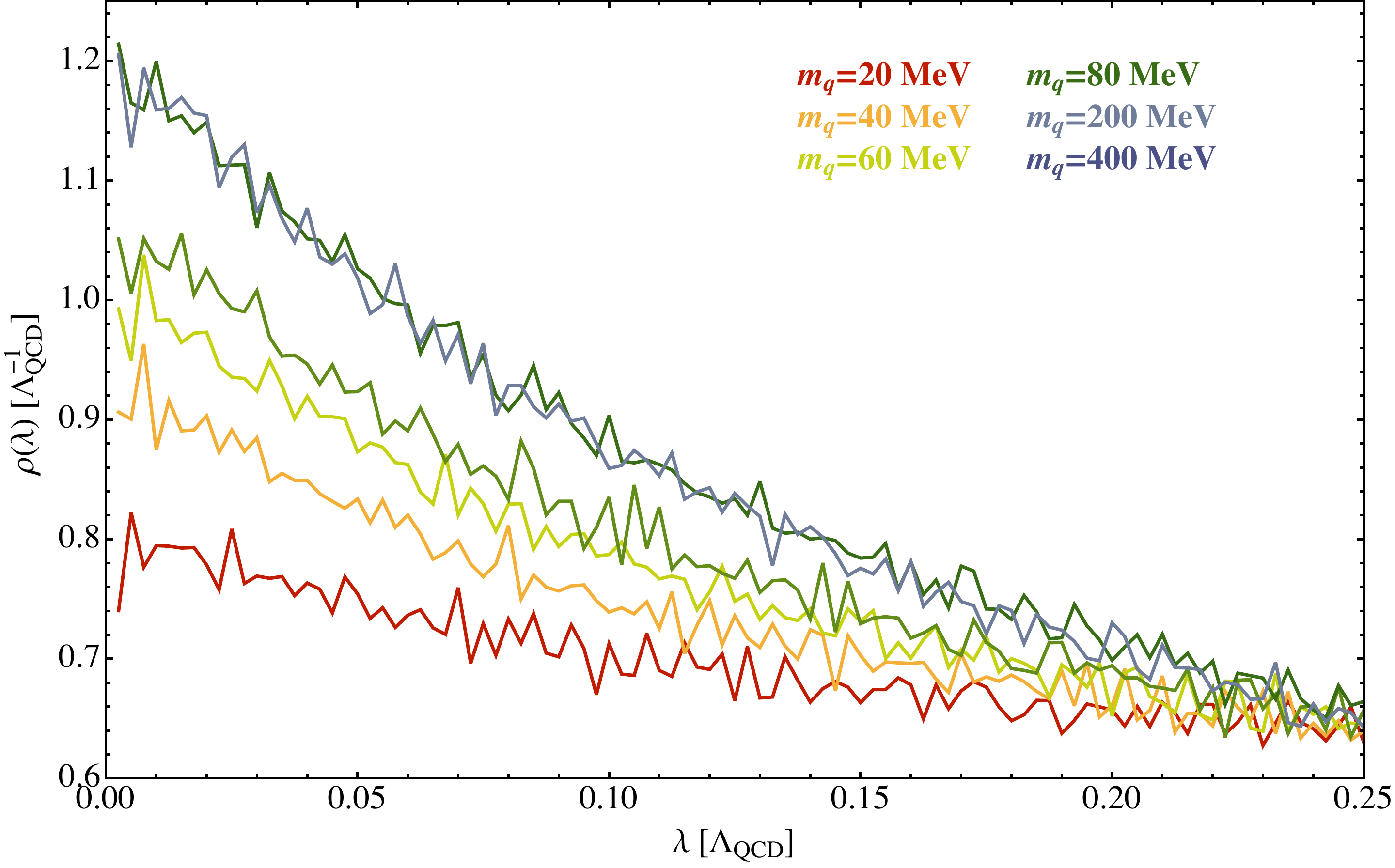}
\caption{Saturation of the Dirac spectrum profile at small $\lambda$ for growing quark mass}
\label{fig:eigfit}
\end{figure}

Other two interesting features of the Instanton Model can be revealed from Fig~\ref{fig:eigfit}: (i) for $m_q\lesssim m^\star$ the density of quasi zero-modes increases as the quark mass decreases  and (ii) for $m_q\gtrsim m^\star$ the near zero-mode part of the spectrum becomes practically independent on the quark mass. These two non-trivial dynamical effects are in fact related and can be explained as follows.
Quark loops are known to generate strong non-local correlations between pseudoparticles of opposite topological charge. Such correlations tend to suppress configurations in which one or more pseudoparticles are located far from all others.  Hence, the contribution of the fermionic determinant leads to a  reduction of the density of {\it nearly} zero-modes, as completely isolated instantons are known to have {\it exact} zero-modes.  As the quark mass gets larger, such a topological screening becomes less and less effective and the population of near exact zero-modes increases, explaining  the rise of the peak of eigenvalue density near the origin. On the other hand, we have seen that for $m_q\gtrsim m^\star$ the contribution of the fermionic determinant is suppressed and fermion-induced topological screening completely disappears. As a result, the low-virtuality sector of the Dirac spectrum stops depending on the quark mass. These are a interesting  predictions of the instanton liquid model that can be checked with unquenched lattice calculations.

\section{Conclusion}
In this talk we have used the IILM as a tool to investigate the dynamical mechanisms which drive the dynamics and the structure of hadrons, in the light-quark sector of QCD. We have shown that the model's predictions are consistent with QCD in the pion mass regime $\lesssim 500~$MeV, where we expect chiral dynamics to play an important role. The model reproduces $\chi$pt prediction for the spectrum of the Dirac operator and is consistent with the lattice data for pion and nucleon masses in the range $m_{\pi}\approx300-600$ MeV. 

Having checked that chiral dynamics is correctly encoded in the instanton model, we have exploited our analytic handling on instanton-induced correlations  to study the dynamical mechanisms involved in the transition into the chiral regime, at a microscopic level. We have identified a mass scale  $m^\star=80$~MeV above which we do not expect QCD correlators to be dominated by chiral dynamics and above which quenched calculations should begin to approximate full QCD.

\end{document}